\documentclass[aps,prl,twocolumn,showpacs]{revtex4-1}

\usepackage{graphicx}% Include figure files
\usepackage{dcolumn}% Align table columns on decimal point
\usepackage{bm}% bold math
\usepackage{hyperref}% add hypertext capabilities
\usepackage[mathlines]{lineno}% Enable numbering of text and display math
\usepackage{amsmath,latexsym}
\usepackage{color, epstopdf, float}

\begin{document}

\title{Giant asymmetric self-phase modulation in superconductor thin films}

\author{Charles Robson}
\author{Fabio Biancalana}
\affiliation{School of Engineering and Physical Sciences, Heriot-Watt University, EH14 4AS Edinburgh, UK \\}

\date{\today}

\begin{abstract}
Self-phase modulation (SPM) of light pulses is found to occur strongly, at low incident intensities, in the coupling of light with superconductors. We develop a theory from a synthesis of the time-dependent Ginzburg-Landau (TDGL) equation and basic electrodynamics which shows the strongly non-linear phase accumulated in the interaction. Unusually, the SPM of the pulse in this system is found to be highly asymmetric, producing a strongly redshifted spectrum when interacting with a superconducting thin film, and it develops in just a few nanometers of propagation. In this paper we present theoretical results and simulations in the THz regime, for both hyperbolic secant and supergaussian-shaped pulses.
\end{abstract}

\maketitle

Self-phase modulation is one of the most basic and best-known nonlinear effects in the coupling of light with matter. It has been extensively studied since its discovery in 1967 \cite{Shimizu} and is due to a pulse of light inducing a change in refractive index in the material in which it propagates, which in turn affects the pulse itself by altering its spectrum; the pulse spectrum can broaden significantly and this broadening has found applications in optical technologies (e.g. fast optical switching, wavelength conversion, and many others) \cite{Agrawal1}.

In this paper we present for the first time theory and simulations describing the SPM of a light pulse interacting with a superconductor. This interacting system is found to be highly nonlinear, exhibiting substantial frequency generation over very small (nanometre) length scales -- we hope this work will generate further research into the field of highly nonlinear optics in superconducting systems.

Superconductivity is a state of matter in which electrical resistance vanishes and external magnetic fields can be expelled (depending on whether the superconductor is itself type-I or type-II, \cite{Annett}).  Historically, these systems have been studied extensively \cite{Annett,Grosso}: one of the most powerful approaches being the Ginzburg-Landau theory \cite{Landau1}, a phenomenological model which works very well in describing the dynamics of the superconductor's order parameter $\psi$, that has been shown to emerge from the microscopic BCS theory in certain limits \cite{BCS,Gorkov1}. The interaction between time-varying fields and superconductors has been studied for several decades and is an active field \cite{Eliashberg1,Eliashberg2,Testardi,Kemper,Smallwood,Sobota,Sentef}, however the interplay between superconductivity and ultrafast nonlinear optics is relatively new and is a field that is emerging in the last few years \cite{Matsunaga,lara,first_paper}, due to its potentially ground-breaking applications. In this paper we utilise the time-dependent Ginzburg-Landau  (TDGL)  equation \cite{Tang} (setting the electric scalar potential to zero; see below for a discussion of this choice):
\begin{equation} \label{eq:TDGL}
\frac{\hbar^2}{2m^*D}\partial_{t}\psi + \frac{1}{2m^*}\left( \mathbf{p}-\frac{q}{c}\mathbf{A} \right)^2 \psi + \alpha\psi + \beta|\psi|^2\psi=0,
\end{equation}
where $\bf{p}$ is the electron momentum operator, $\psi=\sqrt{2n_c} e^{i\phi}$ is the complex order parameter describing the superconductor thermodynamical state with phase $\phi$ and Cooper pair density $n_{c}$, $\mathbf{A}$ is the pulse vector potential, $m^*$ is the mass of a Cooper pair (approximately equal to twice the free electron mass, neglecting the binding energy), $q=-2e$ is the Cooper pair charge (equal to twice the free electron charge $-e$), $D$ is a diffusion constant, $\alpha$ and $\beta$ are phenomenological parameters having units of energy and energy$\times$volume respectively and are explicitly given by $\alpha=\alpha_{0}\left(T-T_{c}\right)$, $\beta=\beta_{0}$, where $\alpha_{0}$ and $\beta_{0}$ are constants specific for the material used. $T$ is the temperature of the superconducting system and $T_{c}$ is the critical temperature below which superconductivity emerges.

We make the simplifying assumption that the electric field of the incident light (and thus its vector potential) is polarized along the $x$-direction only, i.e. $\mathbf{A}=[A(x,y,z,t),0,0]$. Expanding equation (\ref{eq:TDGL}) gives:
\begin{equation} \label{eq:general}
\begin{split}
\frac{\hbar^2}{2m^* D}\partial_t\psi - \frac{\hbar^2}{2m^*}{\nabla^2}\psi +& \frac{i\hbar q}{2m^*c}(\nabla \cdot \mathbf{A})\psi + \frac{i\hbar qA}{m^*c}(\nabla \psi) \\ & + \frac{q^2 A^2}{2m^*c^2}\psi + \beta|\psi|^2 \psi + \alpha\psi=0.
\end{split}
\end{equation}
The three equations on which our theory is based are:
\begin{equation} \label{eq:TDGL2}
\frac{\hbar^2}{2m^* D}\partial_{t}\psi + \frac{q^2 A^2}{2m^* c^2}\psi + \alpha\psi + \beta|\psi|^2 \psi = 0,
\end{equation}
\begin{equation} \label{eq:boxA}
\Box A = \left( \frac{1}{c^2}\partial_{t}^2 - \nabla^2 \right) A = \frac{1}{\epsilon_0 c}J_s,
\end{equation}
\begin{equation} \label{eq:supercurrent}
J_s = -\frac{q^2}{m^* c}|\psi|^2 A;
\end{equation}

We derive the two coupled equations describing the interacting system from TDGL theory through equation (\ref{eq:TDGL2}) (which directly follows from (\ref{eq:general}) where spatial dynamics have been disregarded as they are negligible and $\nabla \cdot \mathbf{A}=0$, see below for rationale), and from simple electrodynamics, i.e. equation (\ref{eq:boxA}). Equation (\ref{eq:boxA}) is a standard electrodynamical expression for the dynamics of the vector potential which is commonly employed as it holds in the Lorenz gauge; in this paper however we work in a different gauge: the phason gauge, see Ref. \cite{Koyama} for details. Equation (\ref{eq:boxA}) still holds in the phason gauge because $\nabla \cdot \mathbf{A}$ vanishes,  assuming charge neutrality \cite{Berm} (scalar potential therefore is taken as zero) in the superconducting film.

Equation (\ref{eq:supercurrent}) \cite{London,Tang} gives the supercurrent $J_s$ induced by the optical field (a phase gradient term contributing to the supercurrent has been disregarded due to its small magnitude).

Using the slowly varying envelope approximation (SVEA) to describe the incident light pulse (and making a Galilean transformation into the frame of the travelling pulse) results in the following two equations describing the coupling of light and superconductor:
\begin{equation} \label{eq:new1}
\frac{\hbar^2}{2m^* D}\frac{d\psi}{dt} + \frac{q^2 |\varepsilon|^2}{4m^* \omega_{0}^2}\psi + \alpha\psi + \beta|\psi|^2 \psi = 0,
\end{equation}
\begin{equation} \label{eq:new2}
i\partial_{z}\varepsilon - \frac{q^2 |\psi|^2}{2\epsilon_0 \omega_0 m^* c}\varepsilon = 0.
\end{equation}
In Eq. (\ref{eq:new2}), $\omega_0$ is the carrier frequency of the pulse and $\varepsilon$ is the envelope of its electric field defined by: $E = \frac{1}{2}\left[ \varepsilon e^{-i(\omega_0 t - k_0 z)} + \varepsilon^* e^{i(\omega_0 t - k_0 z)} \right]$, where $k_{0}\equiv\omega_{0}/c$. Equations (\ref{eq:new1}) and (\ref{eq:new2}) contain all the information required to study SPM in this interacting system.

Equation (\ref{eq:new2}) is of the same form as that describing the propagation of a pulse through a fiber {\cite{Agrawal1}} \emph{except} in the case of the fiber the cubic term goes like $|\varepsilon|^2 \varepsilon$ whereas here we have a $|\psi|^2 \varepsilon$ term instead. The nonlinear interplay between the superconductor (its state represented by $\psi$) and the pulse is the cause of the pulse evolution and spectral broadening. Note that in Eq. (\ref{eq:new2}) we have neglected the dispersion of the medium, which for a relatively long pulse can be discarded, analogously to what is done for optical fibres {\cite{Agrawal1}} .

\begin{figure}
\centering
\includegraphics[width=6cm]{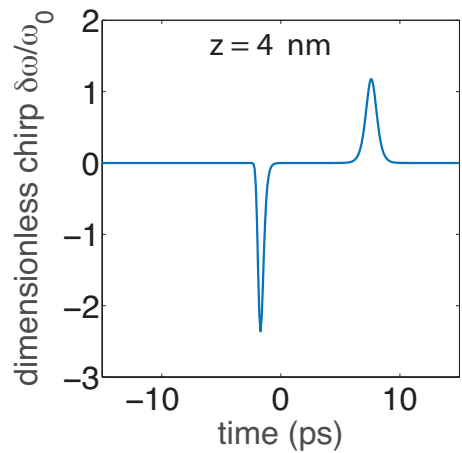}
\caption{Example of highly asymmetric chirp across pulse width caused by nonlinear light-superconductor interplay (length of propagation $z$ shown at top). See Agrawal \cite{Agrawal1} for the corresponding figure of SPM in a fiber in which there are two peaks present (as seen here) but the chirp is strongly asymmetric with respect to time. The asymmetry shown here is the root cause of the strongly redshifted spectrum discussed in this paper.}
\label{fig:chirp1}
\end{figure}

Our numerical results show that the pulse spectrum broadens whilst its intensity profile stays constant, as in standard SPM \cite{Agrawal1}, however the spectrum broadens asymmetrically towards redshifted frequencies. The physical reason behind this asymmetry can be understood as components of the pulse losing energy as they continually interact with the superconductor, giving energy to and locally destroying Cooper pairs. The nonlinear phase shift is found to be, by using Eqs. (\ref{eq:new1}-\ref{eq:new2}), $\theta_{NL}(t) = - q^2 |\psi(t)|^2 z / (2\epsilon_0 m^* c \omega_0)$. From the chirp expression $\delta\omega(t)=-\partial \theta_{NL}(t)/\partial t$ \cite{Agrawal1} it follows that $\delta\omega(t)=( q^2 z / 2\epsilon_0 m^* c \omega_0 )\partial |\psi(t)|^2/\partial t$. An example of a result from our simulations showing the chirp induced on a supergaussian-shaped incident THz pulse is given in Fig. {\ref{fig:chirp1}}. Note that this chirp develops after only a few nanometers of propagation, due to the very large nonlinearity of superconductors.

The asymmetry of the order parameter with respect to time is more pronounced when the order parameter drops sharply, for example under the influence of a supergaussian pulse, and therefore the spectrum is broadened to a greater extent than under the influence of a less sharp incident pulse, such as the hyperbolic secant. In the well-known case of standard SPM in an optical fiber \cite{Agrawal1} the spectral broadening is symmetric because it is contingent on the symmetrical time variation of the \emph{pulse intensity} as opposed to the time variation of the Cooper pair number density $|\psi|^2$ as is the case here. This is the novel result of this paper.

\begin{figure}[htbp]
\centering
\includegraphics[width=9cm]{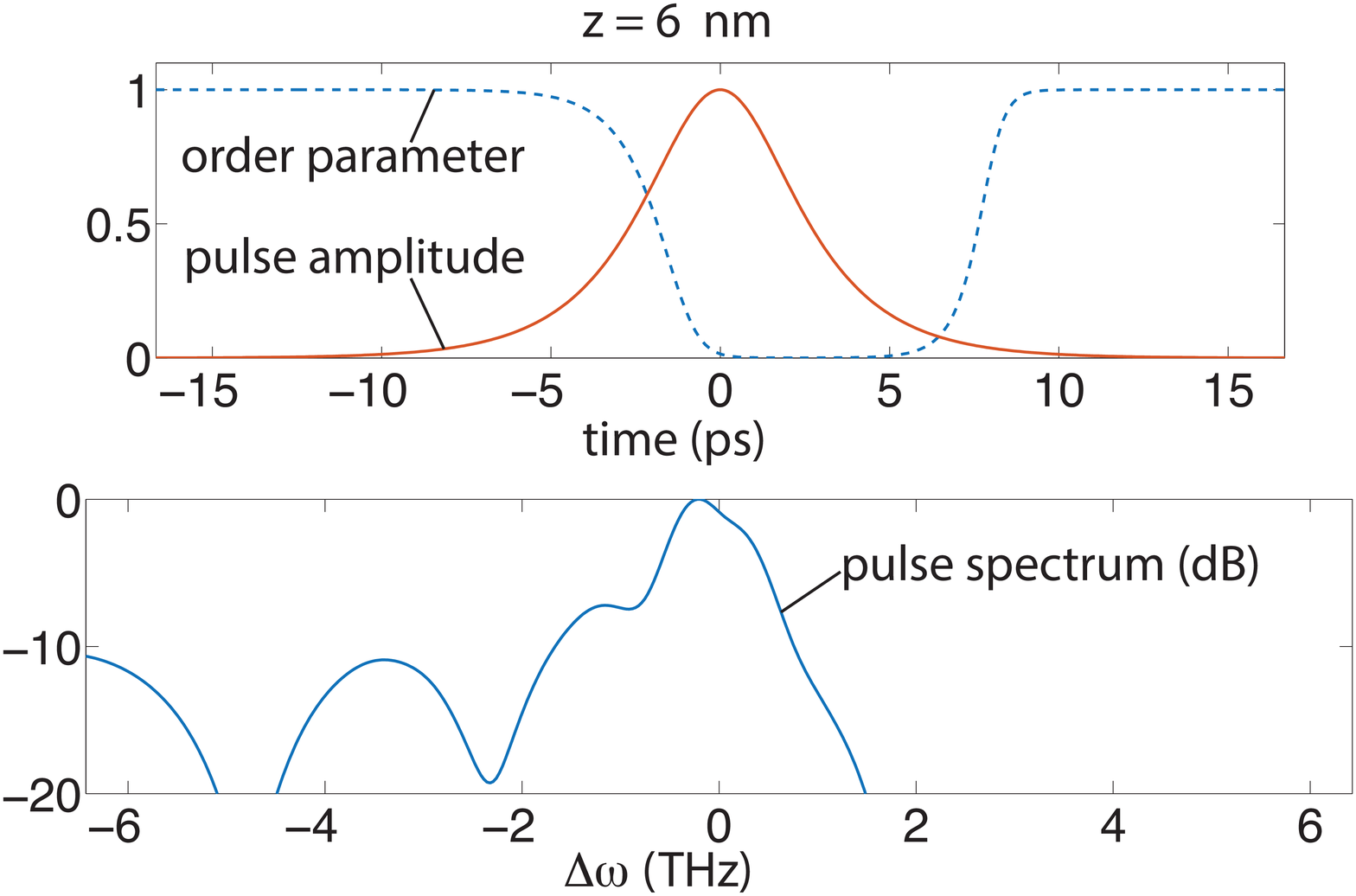}
\caption{Results for incident sech-shaped pulse in the THz regime. Top: order parameter (dashed) and pulse (solid) in the comoving frame (both normalised to one); Bottom: widened spectrum of pulse (in decibels).}
\label{fig:diptych1}
\end{figure}

\begin{figure}[htbp]
\centering
\includegraphics[width=10cm]{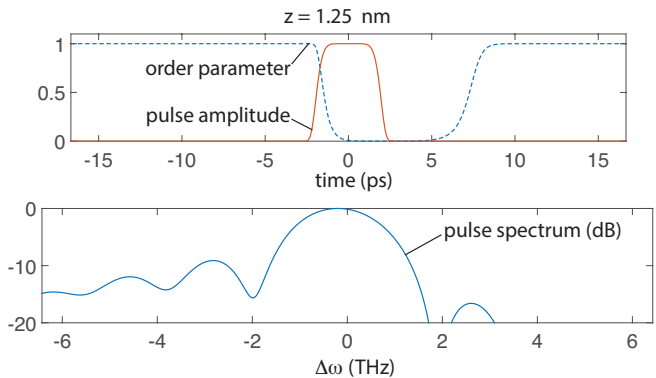}
\caption{Results for incident supergaussian-shaped pulse in the THz regime. Top: order parameter (dashed) and pulse (solid) in the comoving frame (both normalised to one); Bottom: widened spectrum of pulse (in decibels).}
\label{fig:diptych2}
\end{figure}

Here we present the results from our numerical solutions of equations (\ref{eq:new1}) and (\ref{eq:new2}) in the THz regime, which is of interest in applications \cite{Matsunaga,lara}. However, nothing prevents the application of our model to different frequency ranges, for instance in the near-IR or the visible, although in these regions the nonlinearity drops sharply, since it depends on a factor $\omega^{6}$, while in the THz regime the nonlinear susceptibility of superconductors is the largest for all known materials \cite{first_paper}.

We simulate light of central frequency $\omega_{0} = 15$ THz, for two different pulse shapes: a hyperbolic secant $\varepsilon\sim\rm{sech}(t/t_0)$ and a sharper supergaussian $\varepsilon\sim\rm{exp}(-(t/t_0)^8)$. We use the following parameters throughout: niobium type-II superconductor at temperature $T=4.2$K, pulse widths $T_{\rm{sech}}= 5.3\rm{ps}$, $T_{\rm{supergaussian}}= 3.8\rm{ps}$, and an intensity peak for both of $I=36 \ \rm{kW}/\rm{cm}^2$ (a very low intensity easily achievable experimentally). We choose to study niobium as it has the highest critical temperature ($T_{c}=9.25\rm{K}$) of any elemental superconductor \cite{Piel,Stromberg}. For the temperature at which we simulate the solutions to the equations ($T=4.2$K) niobium has an energy gap of $2\Delta(T=4.2K)/\hbar= 5$ THz; we work under the constraint that our incident light pulse must have greater energy than this so that the TDGL theory is applicable. The spatial scale over which the SPM takes place can be calculated from Eqs. (\ref{eq:new1}-\ref{eq:new2}), and is $z_{\rm SPM}=2\beta\epsilon_{0}m^{*}c\omega_{0}/(q^2|\alpha(T)|)$, which is at the selected temperature equal to $0.28$ nm. This is an important result: due to its massive nonlinearity in the THz regime, light propagation through a thin film of superconductor provides a large SPM phase shift.

As seen in Fig. \ref{fig:diptych1}, the sech-shaped pulse (maintaining its shape throughout propagation) causes the order parameter to dip and then recover as it interacts with the superconductor (this has been studied in more depth in our previous work \cite{first_paper}) whilst its spectrum redshifts substantially, producing lateral lobes. The supergaussian pulse simulation result shown in Fig. \ref{fig:diptych2} indicates a large redshift over a \emph{shorter} pulse propagation distance, this is because the sharper pulse front creates more asymmetrical order parameter dynamics: the order parameter drops sharply, whereas the sech pulse destroys the Cooper pairs more gradually, this can also be understood from the chirp formula derived above. The nonlinear effects (the redshifting of the spectrum) are stronger as incident frequencies decrease; we have studied this inverse proportionality between frequency and nonlinearity in detail in Ref. \cite{first_paper}. In particular, note that the scale of propagation for observing the SPM lobes is of the order of nanometers, as opposed to optical fibres, where the scale can be of the order of hundreds of meters \cite{Agrawal1}.

An induced nonlinear phase shift is a standard feature of SPM, however, an interesting and novel feature of the system studied here is that the phase shift mirrors and encodes the order parameter dynamics (as can be seen in our formula for $\theta_{NL}(t)$ above) whilst growing with propagation length $z$. In standard fiber optic SPM the phase shift across the pulse depends on its intensity profile whereas here the interacting system enforces an order parameter-dependent phase. We believe that measuring this phase shift across the interacting pulse may prove to be a useful diagnostic tool in future for studying the ultrafast dynamics of a superconductor's order parameter, and therefore Cooper pair density. The maximum nonlinear phase shift of the pulse roughly follows $\theta_{max} \approx \left( 2M - 1\right) \pi$, see also Ref. \cite{svelto}, where $M$ is the number of peaks in the asymmetrically widened pulse spectrum.

In conclusion, in this paper we have presented the first exposition of self-phase modulation of a pulse interacting with a superconducting thin film. The theory was formed using only basic TDGL theory and electrodynamics. Simulations have shown a giant optical nonlinearity and spectral redshift for very low incident intensities and very short propagation distances of the order of nanometers. We hope this will stimulate further work in the burgeoning field of ultrafast superconducting nonlinear photonics.

% \section{References}

% [Full references (to aid the editor and reviewers) must be included as well on a fifth informational page that will not count against page length.]

% Bibliography
% \bibliography{sample}

% Full bibliography added automatically for Optics Letters submissions; the following line will simply be ignored if submitting to other journals.
% Note that this extra page will not count against page length
% \bibliographyfullrefs{sample}
 
%Manual citation list

\end{document}